# Evidence for the Role of Instantons in Hadron Structure from Lattice QCD

M.-C. Chu,[a] J. M. Grandy,[b] S. Huang,[c,d] and J. W. Negele[d] *

[a]W. K. Kellogg Radiation Laboratory, Caltech, 106–38 Pasadena, California 91125

[b]T-8 Group, MS B-285, Los Alamos National Laboratory, Los Alamos, New Mexico 87545

[c]Department of Physics, FM-15, University of Washington, Seattle, Washington 98195

[d]Center for Theoretical Physics, Laboratory for Nuclear Science, and Department of Physics
Massachusetts Institute of Technology, Cambridge, Massachusetts 02139

Cooling is used as a filter on a set of gluon fields sampling the Wilson action to selectively remove essentially all fluctuations of the gluon field except for the instantons. The close agreement between quenched lattice QCD results with cooled and uncooled configurations for vacuum correlation functions of hadronic currents and for density-density correlation functions in hadronic bound states provides strong evidence for the dominant role of instantons in determining light hadron structure and quark propagation in the QCD vacuum.

## 1. Introduction

To elucidate the role of instantons in hadron structure, we consider correlation functions which characterize the gross structure of hadrons and quark propagation in the QCD vacuum and which are well described by quenched lattice QCD calculations which sample the full Wilson action. These calculations include all the fluctuations and topological excitations of the gluon field and thus include the full perturbative and non-perturbative effects of the short range Coulomb and hyperfine interactions, confinement, and instantons. We then use cooling to remove essentially all fluctuations of the gluon field except for the instantons which, because of their topology, cannot be removed by local minimization of the action. Thus, both the Coulomb interaction and confinement are almost completely removed while retaining most of the instanton content. To the extent to which the gross features of light hadron structure and quark propagation in the QCD vacuum are unaffected by removing all the gluonic modes except instantons, we have strong evidence for the dominant role of instantons.

The vacuum correlation functions we consider are the point-to-point equal time correlation functions of hadronic currents

$$R(x) = \langle \, \Omega \, | \, T \, J(x) \, \bar{J}(0) \, | \, \Omega \, \rangle$$

discussed in detail by Shuryak [1] and recently calculated in quenched lattice QCD [2].

To characterize the gross properties of hadrons, in addition to the mass, we consider quark density-density correlation functions [3–5] $\rho(x) = \langle h|\hat{\rho}(x)\,\hat{\rho}(0)|h\rangle$. In contrast to wave functions, which have large contributions from the gluon wave functional associated with the gauge choice or definition of a gauge invariant amplitude [6], the density-density correlation function is a gauge-invariant physical observable which directly specifies the spatial distribution of quarks.

This work is strongly motivated by the physical arguments and instanton models of QCD [7–9] in which the zero modes associated with instantons produce localized quark states and quark propagation takes place primarily by hopping between these states.

*Based on talks presented by S. Huang and J. W. Negele. This work is supported in part by funds provided by the U. S. Department of Energy (D.O.E.) under contracts #DE-AC02-76ER03069 and #DE-FG06-88ER40427, and the National Science Foundation under grant #PHY 88-17296.

## 2. Lattice Calculations

Cooling [10, 11] is used as a filter to extract the instanton content of 19 gluon configurations obtained by sampling the standard Wilson action on a $16^3 \times 24$ lattice at $\frac{6}{g^2} = 5.7$. We used the Cabibbo-Marinari [12] algorithm with three $SU(2)$ subgroups and $\beta = \infty$ to minimize the action sequentially on each link of the lattice in each cooling step.

To monitor the filtering of different degrees of freedom as a function of cooling steps, we measure several gluonic observables. Short wavelength fluctuations giving rise to the Coulomb and hyperfine interactions are reflected in the total action $S$. Confinement is monitored by measuring the string tension extracted from a $4 \times 7$ Wilson loop, and we refer to results from an earlier calculation at the same value of $\beta$ with the same cooling algorithm [13]. Finally, to monitor the instanton content, we measure the topological charge $\langle Q \rangle$ and topological susceptibility $\langle Q^2 \rangle$, using the simplest expression for the topological charge density $Q(x_n) = -\frac{1}{32\pi^2} \epsilon_{\mu\nu\rho\sigma} \, \mathrm{ReTr}\, [U_{\mu\nu}(x_n) U_{\rho\sigma}(x_n)]$. Note that for a random ensemble of Poisson distributed instantons and anti-instantons, $\langle Q \rangle = 0$ and $\langle Q^2 \rangle = I + A$, the number of instantons plus anti-instantons.

The correlation functions we calculate in the pseudoscalar, vector, nucleon and Delta channels are

$$R(x) = \langle \Omega | T\, J^p(x) \bar{J}^p(0) | \Omega \rangle \ ,$$
$$R(x) = \langle \Omega | T\, J_\mu(x) \bar{J}_\mu(0) | \Omega \rangle \ ,$$
$$R(x) = \tfrac{1}{4} \mathrm{Tr}\, \left( \langle \Omega | T\, J^N(x) \bar{J}^N(0) | \Omega \rangle\, x_\nu\, \gamma_\nu \right) \ ,$$
and
$$R(x) = \tfrac{1}{4} \mathrm{Tr}\, \left( \langle \Omega | T\, J^\Delta_\mu(x) \bar{J}^\Delta_\mu(0) | \Omega \rangle\, x_\nu\, \gamma_\nu \right) \ ,$$

where

$$J^p = \bar{u}\, \gamma_5\, d \ ,$$
$$J_\mu = \bar{u}\, \gamma_\mu\, \gamma_5\, d \ ,$$
$$J^N = \epsilon_{abc} [u^a\, C\, \gamma_\mu\, u^b]\, \gamma_\mu\, \gamma_5\, d^c \ ,$$
and
$$J^\Delta_\mu = \epsilon_{abc} [u^a\, C\, \gamma_\mu\, u^b]\, u^c \ .$$

As in Refs. [1] and [2], we consider the ratio of the correlation function in QCD to the correlation function for non-interacting massless quarks, $\frac{R(x)}{R_0(x)}$, which approaches one as $x \to 0$ and displays a broad range of non-perturbative effects for $x$ of the order of 1 fm. The effects of lattice anisotropy are removed by calculating $R_0(x)$ on the same lattice as $R(x)$ and measuring the ratio for a cone of lattice sites concentrated around the diagonal. Finite lattice volume effects are corrected by subtracting the contributions of first images, and the correlation functions are fit by a spectral function parameterized by a resonance mass, the coupling to the resonance, and the continuum threshold. Hadron density correlation functions, $\langle h | \rho_u(x) \rho_d(0) | h \rangle$, are calculated as in Refs. [4], [5], and [14] for the pion, rho and nucleon, where $\rho_u = \bar{u} \gamma_0 u$ and similarly for $\rho_d$. Image corrections for finite volume effects are applied as in Ref. [14].

A significant conceptual issue in comparing observables calculated using cooled configurations with uncooled results is how to change the renormalization of the bare mass and coupling constant as the gluon configurations are cooled. We use the physical pion and nucleon masses to determine $\kappa$ and $a$ for the cooled configurations, with the result that $a$ changes by $\sim 16\%$ after 25 cooling steps and the rho mass remains unchanged within errors with this value of $a$.

## 3. Instanton content of the gluon vacuum

To provide a clear picture of how cooling extracts the instanton content of a thermalized gluonic configuration, we display in Fig. 1 the action density $S(1, 1, z, t)$ and topological charge density $Q(1, 1, z, t)$ for a typical slice of a gluon configuration before cooling and after 25 and 50 cooling steps. As one can see, there is no recognizable structure before cooling. Large, short wavelength fluctuations of the order of the lattice spacing dominate both the action and topological charge density. After 25 cooling steps, three instantons and two anti-instantons can be identified clearly. The action density peaks are completely correlated in position and shape with the topological charge density peaks for instantons and with the topological charge density valleys for anti-instantons. Note that both the action and topological charge densities are reduced by more than two orders of magnitude, so that the fluctuations removed by cooling are several orders of magnitude larger than the topological excitations that are retained. Further cooling to 50 steps results in the annihilation of the nearby instanton–anti-instanton pair but retains the well separated instantons and anti-instanton.

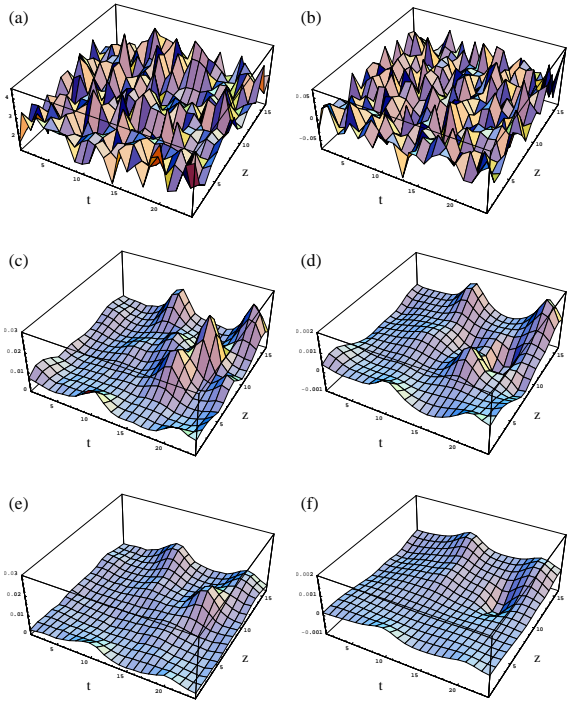

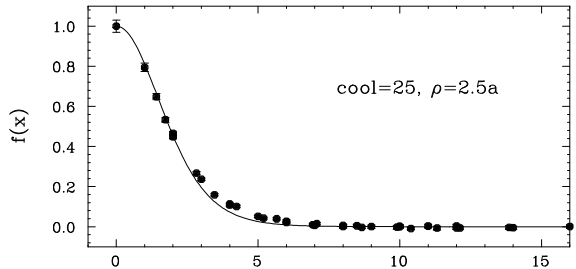

Figure 2. Topological charge density–density correlation function after 25 cooling steps. Lattice measurements are denoted by solid points with error bars. The curve shows the best fit obtained using a convolution of the topological charge density for a single instanton size $\rho$.

Figure 1. Cooling history for a typical slice of a gluon configuration at fixed $x$ and $y$ as a function of $z$ and $t$. The left column shows the action density $S(1,1,z,t)$ before cooling (a), after cooling for 25 steps (c) and after 50 steps (e). The right column shows the topological charge density $Q(1,1,z,t)$ before cooling (b), after cooling for 25 steps (d) and after 50 steps (f).

As expected, the action is dominated by the short range modes and is therefore very strongly affected by cooling. Denoting the action of a single instanton by $S_0 = 8\pi^2/g^2$, $<S>/S_0$ decreases from 20,211 before cooling to 64 and 31 at 25 and 50 sweeps respectively. The topological charge is less sensitive to short range modes, and

$Q^2$ is essentially constant at $\sim 25 \pm 10$ throughout the cooling. The string tension in lattice units, $\sigma a^2$, is $0.18, 0.05$, and $0.03$ at $0, 25,$ and $50$ cooling steps. At cooling step 25, the action and string tension have dropped to 0.3% and 27% of the uncooled values, indicating a dramatic reduction in perturbative and confining effects. The difference between $<S>/S_0 \sim 65$ and $\langle Q^2 \rangle \sim 25$ indicates that there are sufficient nearby instanton–anti-instanton pairs in each configuration that we have not yet reached the dilute regime where $\langle Q^2 \rangle \sim A + I$, which only sets in beyond 50 steps. We regard the configurations cooled with 25 steps as providing a more complete description of the instanton content of the original configurations, and will therefore emphasize them in our subsequent calculation of hadronic properties.

To estimate the instanton size we measure the topological charge density correlation function $f(x) = \sum_y Q(y)Q(x+y)$ where $Q(y)$ is the topological charge density at point $y$ and the sum is over the whole lattice. The ensemble average of $f(x)$ at cooling step 25 is displayed in Fig.2. The strong peak at small $x$ is the correlation of a single instanton or anti-instanton with itself. The vanishing of $\langle f(x) \rangle$ at large $x$ implies that the topological charge is uncorrelated at this larger distance and thus averages to zero.

If we assume that all instantons are well separated, we would expect that each individual peak can be approximated by the analytic instanton topological charge density $Q_\rho(x) = \frac{6}{\pi^2 \rho^4}\left(\frac{\rho^2}{x^2+\rho^2}\right)^4$ where $\rho$ is the size parameter. Although in principle one should fit with a distribution of values of $\rho$, a first approximation is obtained by using a single value of $\rho$ which we will interpret as an average value. A convolution of $Q_\rho(x)$ with itself defines a function which can be used to fit the lattice data with $\rho$ as the fitting parameter. The continuous curve in Fig. 3 is the fitted result with $\rho = 2.5a$ for 25 cooling steps and we similarly obtain $\rho = 2.8a$ for 50 cooling steps. The fit fails to reproduce the detailed shape for $x/a \sim 5$, both because of the range in instanton sizes and the nonlinear overlap of instantons as observed in Fig.1. By analyzing the sizes of lumps in the topological charge density, we also obtain rough histograms of the distribution of instantons with $\rho$ which are consistent with the average values quoted above.

To obtain results in physical units, we use the scale $a$ determined from the nucleon mass, which decreases from 0.168 fm before cooling to 0.142 fm and 0.124 fm at 25 and 50 cooling steps respectively. The 16% decrease in $a$ after 25 cooling steps is quite modest, so that our qualitative conclusions are relatively insensitive to the scale change. The values at 25 (50) cooling steps for the instanton size of 0.36 fm (0.35 fm), for the instanton density of 1.64 fm$^{-4}$ (1.33 fm$^{-4}$), and for topological charge susceptibility, $\chi$, of [177 MeV]$^4$ ([200 MeV]$^4$) compare well with the values 0.33 fm, 1.0 fm$^{-4}$, and [180 MeV]$^4$ used in instanton models by Shuryak and collaborators [8].

A similar analysis of cooled configurations has previously been carried out for SU(2) with smaller lattices and slightly different techniques [15], where the positions and magnitudes of peaks in $S(x,y,z,t)$ were used to determine the distribution of sizes of instantons.

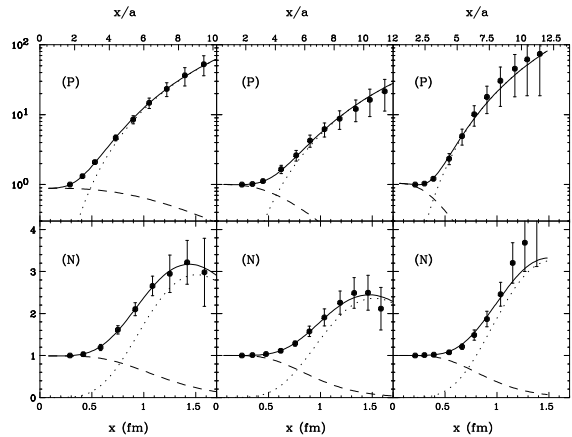

Figure 3. Comparison of uncooled and cooled vacuum correlation function ratios, $\frac{R(x)}{R_0(x)}$, for pseudoscalar currents ($P$) and nucleon currents ($N$). The left, center, and right panels show results for uncooled QCD, 25 cooling steps, and 50 cooling steps respectively. The solid points with error bars denote lattice correlation functions extrapolated to $m_\pi = 140$ MeV. The solid lines denote fits to the correlation functions using a three-parameter spectral function, and the dashed and dotted curves show the contributions of the continuum and resonance components of the spectral functions respectively. The upper scale shows the spatial separation in lattice units and the lower scale shows the separation in physical units.

## 4. Hadronic observables in the cooled vacuum

In the top panels of Fig. 3, we show the ratio of vacuum correlation functions for interacting to non-interacting quarks $\frac{R(x)}{R_0(x)}$ in the pseudoscalar channel for uncooled QCD, 25 cooling steps, and 50 cooling steps. This channel is by far the most attractive of all the meson channels, as reflected in the fact that the correlation function for interacting quarks is roughly 50 times larger than for free quarks, and is thus the only channel to be plotted on a log scale. Since the pion mass is used to determine the base quark mass, masses of the pion resonance term in Fig. 3 are constrained to be fixed at 140 MeV. Note that after 25 cooling steps, the correlation function is qualitatively similar to the uncooled result, although the magnitude at 1.5 fm is roughly half as large. After an additional 25 cooling steps, the peak grows in strength. Apparently, although the distribution of instantons after 50 steps is more dilute and less representative of the QCD vacuum than after 25 steps, it reproduces the uncooled correlation function slightly better. To assure that this behavior is not a statistical artifact, in this and every other channel we analyzed two independent sets of 9 and 10 configurations separately and verified that the same behavior occurred in both cases.

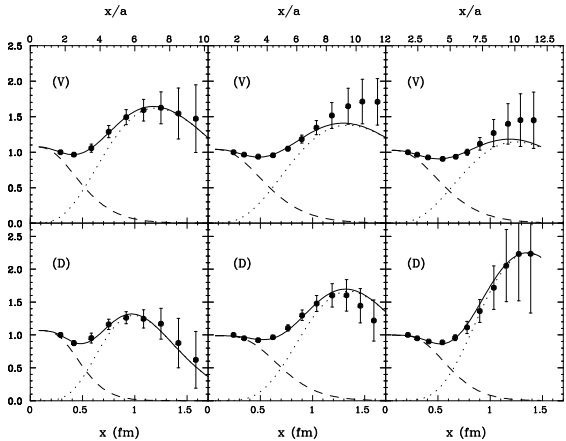

Figure 4. Comparison of uncooled and cooled vacuum correlation function ratios, $\frac{R(x)}{R_0(x)}$, for vector currents ($V$) and Delta currents ($D$). The notation is the same as in Fig. 3.

Analogous results for $\frac{R(x)}{R_0(x)}$ in the nucleon channel are shown in the bottom panels of Fig. 3, where again the nucleon mass is constrained to be constant because it is used to determine the lattice spacing. The behavior is similar to that in the pseudoscalar channel. After 25 sweeps, the correlation function is qualitatively similar to the uncooled result. In detail, the peak also appears lower after cooling, although this time it agrees within errors. After an additional 25 sweeps the peak height increases again, agreeing even more closely with the uncooled result.

The ratios of correlation functions $\frac{R(x)}{R_0(x)}$ for the vector channel are shown in the upper panels of Fig. 4. The $\rho$ mass governing the resonance peak is unconstrained, but does not change significantly with cooling. Furthermore, there is virtually no change in the correlation function ratio with cooling. Similarly, in the $\Delta$ channel the peak position does not shift significantly with cooling and while its height appears to increase somewhat, the errors are consistent with its staying constant.

Density-density correlation functions in the ground state of the pion, rho, and nucleon are shown in Fig. 5. The errors for the uncooled results have been suppressed for clarity since they are comparable to those for the cooled results.

The striking result for both the rho and the nucleon is the fact that the spatial distribution of quarks is essentially unaffected by cooling — instantons alone govern the gross structure of these hadrons, as indeed they also governed vacuum correlation functions of hadron currents in these same channels. The only case in which a noticeable change is brought about by cooling is in the short distance behavior of the ground state of the pion. This is understandable since in the physical pion, in addition to instanton induced interactions, there is also a strong attractive hyperfine interaction arising from perturbative QCD which, combined with the $1/r$ interaction, gives rise to the central peak in the density. In the rho, the hyperfine interaction has much less effect, both because it is repulsive and is 3 times weaker.

It is also noteworthy that the cooled density-density correlation functions shown in Fig. 5 for the $\pi$, $\rho$, and nucleon are comparable (within error bars), strongly suggesting that instantons set the overall spatial scale of these hadrons.

In conclusion, the close agreement between hadronic observables with cooled and uncooled configurations provides strong evidence for the dominant role of instantons in determining hadron structure and quark propagation in the QCD vacuum.

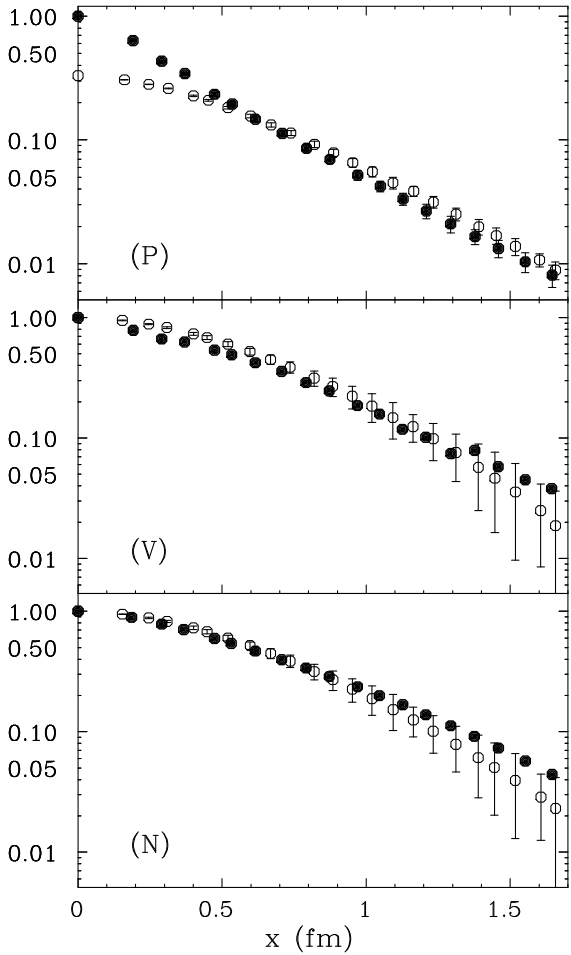

Figure 5. Comparison of uncooled and cooled density-density correlation functions for the pion, rho, and nucleon. The solid circles denote the correlation functions calculated with uncooled QCD, and the open circles with error bars show the results for 25 cooling steps. The rho and pion results are compared for $m_\pi^2 = 0.16$ GeV$^2$ and the nucleon results are compared for $m_\pi^2 = .36$ GeV$^2$. As in Figs. 3 and 4, the separation is shown in physical units. All correlation functions are normalized to 1 at the origin except for the cooled pion correlator, which is normalized to have the same volume integral as the uncooled correlator.